\documentclass[preprint]{aastex}

\def\sqig{$\sim$}

\def\degrees{$^{\circ}$}

\def\source{SMC X-2}
\def\src{SMC X-2}

\begin{document}
\title{The Discovery of an Outburst and
Pulsed X-ray Flux from \src\ from RXTE Observations}

\author{R.H.D. Corbet\altaffilmark{1, 2},
F.E. Marshall\altaffilmark{1},
M.J. Coe\altaffilmark{3},
S. Laycock\altaffilmark{3},
\&
G. Handler\altaffilmark{4}
}

\altaffiltext{1}{Laboratory for High Energy Astrophysics, Code 662,
NASA/Goddard Space Flight Center, Greenbelt, MD 20771}
\altaffiltext{2}{Universities Space Research Association;
corbet@lheamail.gsfc.nasa.gov}
\altaffiltext{3}{Department of Physics \& Astronomy,
University of Southampton, Southampton, SO17 1BJ, U.K.}
\altaffiltext{4}{South African Astronomical Observatory,
P.O. Box 9, Observatory, 7935, South Africa}

\begin{abstract}
Rossi X-ray Timing Explorer All Sky Monitor observations of \src\ show
that the source experienced an outburst in January to April 2000 
reaching a peak luminosity of greater than \sqig10$^{38}$ ergs s$^{-1}$.
RXTE
Proportional
Counter Array observations during this outburst reveal the presence
of pulsations with a 2.37s period.  However, optical photometry of the
optical counterpart showed the source to be still significantly fainter
than it was more than half a year after the outburst in the 1970s when \src\ was discovered.

\end{abstract}
\keywords{stars: individual (\source) --- stars: neutron ---
X-rays: stars}

\section{Introduction}

The first three X-ray point sources to be discovered in the Small
Magellanic Cloud (SMC) were SMC X-1, SMC X-2 and SMC X-3, all of which are
thought to be X-ray binaries. While SMC X-1 is a persistent pulsing source
(e.g. Wojdowski et al. 1998 and references therein),
SMC X-2 and X-3 are transient sources that
were discovered in 1977 with SAS-3 (Clark et
al. 1978) which were both initially very bright 
at a level of \sqig 1.0 $\times$ 10$^{38}$ ergs s$^{-1}$ (2 $-$ 11 keV)
for assumed distances of 71 kpc
and remained in outburst
for approximately one month (Clark, Li, \& van Paradijs 1979). Both sources
were also detected with HEAO-1 A-2 experiment
(Marshall et al. 1979).  However, they have since exhibited little
activity.  SMC X-2 was not detected in an Einstein
IPC survey (Seward \& Mitchell 1980) and was seen in
only
one of two observations made with the ROSAT PSPC detector
(Kahabka \& Pietsch 1996, Haberl \& Sasaki 2000)
at a luminosity of 2.7 $\times$ 10$^{37}$ ergs s$^{-1}$ (0.15 $-$ 2.4 keV).
SMC X-3 was also not detected by Seward \& Mitchell (1980)
but was seen in one ROSAT HRI observation
(Haberl \& Sasaki 2000).
An optical
counterpart for SMC X-2 was proposed by Crampton, Hutchings, \& Cowley
(1978) which was resolved into a pair of stars, ``A'' and ``B'', with the fainter object (``B''),
a main-sequence Be star, identified as the likely counterpart by Murdin,
Morton, \& Thomas (1979).  Infra-red observations
of SMC X-2 during a quiescent period are reported by
Coe et al. (1997). In none of the previous X-ray observations
of SMC X-2 or SMC X-3 were any pulsations detected although they are
expected to be present as most Be star X-ray sources are X-ray pulsars.
Another bright X-ray pulsar candidate within the SMC is the little
studied source H0107-750 which also has a Be star as its proposed
optical counterpart (Whitlock \& Lochner 1994 and references
therein).

Although no other SMC X-ray pulsars apart from SMC X-1
were known for many years there has recently been an explosion
in the discovery of SMC X-ray pulsars. Hughes (1994) discovered
a 2.7s pulsar, and three pulsars were found to be active in
a single RXTE observation of the vicinity of SMC X-3 (Corbet et al. 1998).
Further RXTE, ASCA and SAX observations have now lead to
the identification of 
over 20 X-ray pulsars in the SMC 
(e.g. Haberl \& Sasaki 2000, Yokogawa et al. 2000, and
references therein) and their variability and several optical
identifications suggests that at least the vast majority
are transient Be star systems. The number of known Be
star X-ray binaries in the SMC is thus now starting to rival the
number known in
the Galaxy (e.g. Bildsten et al. 1997).

In this paper we report the detection of activity from the vicinity of
SMC X-2 in early 2000 with the Rossi X-ray Timing Explorer All-Sky Monitor
(ASM). Subsequent observations with the RXTE Proportional Counter Array
(PCA) then revealed the presence of 2.37s pulsations. Contemporaneous
optical photometry of the optical counterpart are compared to the initial discovery photometry. Note that in this paper
we use an assumed distance of
65 kpc to the SMC for consistency with several recent publications
(e.g. Kahabka \& Pietsch 1996, Haberl \& Sasaki 2000)
which is
somewhat smaller than
the 71 kpc adopted by Clark et al. (1978).

\section{Observations and Analysis}

\subsection{RXTE All-Sky Monitor}

The All-Sky Monitor (Levine et al. 1996) on board RXTE consists of
three Scanning Shadow Cameras, sensitive to X-rays in an energy band
of approximately 2-12 keV, which perform sets of 90 second pointed
observations (``dwells") so as to cover \sqig80\% of the sky every \sqig90
minutes. The analysis presented here makes use of daily averaged light
curves constructed from the flux measured in individual dwells. The Crab
produces approximately 75 counts s$^{-1}$ in the ASM over the entire energy range
and ASM observations of blank field regions away from the Galactic center
suggest that background subtraction is accurate to
a systematic level
about 0.1 counts s$^{-1}$ (Remillard \& Levine 1997). The RXTE All Sky Monitor
provides light curves of several hundred sources including \src.

An examination of the light curve obtained from the ASM on \src, shown
in Figure 1, suggested that the source may have increased in brightness
in approximately January 2000. The source was still very faint and
the possible increase in brightness was only visible in a very heavily
smoothed and rebinned version of the ASM light curve.  Because of this
possible increase in flux we requested target of opportunity
observations of the vicinity of
SMC X-2 with the RXTE PCA.

\subsection{Proportional Counter Array}

The Proportional Counter Array (PCA; Jahoda et al. 1996) consists of five
Proportional Counter Units (PCUs) sensitive to photons with
energies between 2 - 60 keV with a total effective area of \sqig6500
cm$^2$. For the entire PCA across the complete energy band the Crab
produces a count rate of about 13,000 counts s$^{-1}$.  The PCA spectral resolution
at 6 keV is approximately 18\% and the collimator gives a field of view
of 1\degrees\ full width half maximum. Due to instrumental
problems not all PCUs are always operated and the observations
reported here use varying numbers of PCUs.

Before the first set of scheduled observations of \src\ was performed
we noted the serendipitous detection of a source consistent with the
location of \src\ during slews over the SMC on 2000 April 9.
In order to localize the position of the X-ray emission from this region
more precisely, and so determine whether it was indeed consistent with 
the location of \src, we
performed additional short slews in right ascension and declination over the position
of \src\ on 2000 April 12 together with short ``stares''.
Additional pointed observations were made between 2000 April 22
21:04 to April 23 02:43 
 and 2000 May 2 (06:09 to 06:48), May 3 (06:05 to 06:47 and 09:34 to 09:58)
and May 5 (16:42 to 17:16).
A log of the observations is given in Table 1. Data
reduction used standard HEAsoft software, and background subtraction
was performed using the model appropriate for faint
sources.

\subsection{Optical}

The optical counterpart of \src\ was observed with the South African
Astronomical Observatory 1.0m telescope on 2000 April 14, two days
after RXTE observation \#2 was performed.  Data were
collected using the SITe4 CCD giving a field of approximately 5 arcminutes
and a detector scale of 0.3 arcseconds per pixel.  Observations were made
through standard Johnson BV and Cousins R$_C$ filters.  The data were reduced using IRAF
and Starlink software, and the instrumental magnitudes were corrected
to the standard system using E region standards. The resulting BVR$_C$
magnitudes are shown in Table 2, and all have an associated
1 $\sigma$ error of $\pm$0.02.

The results may be compared with the photometry of Murdin et al.
(1979) which are also given in Table 2. These authors do not
quote any errors on their results, but given that they only present
their values to one decimal place they could well be $\pm$0.1. Hence
the values for Star A (not the proposed candidate) are essentially in
agreement between the two epochs.

\section{Results}

\subsection{ASM}

The ASM light curve (Fig. 1)
suggests that there was an outburst from
\src\ that occurred
between roughly 2000 January to April. 
However, the faintness of the source makes it
difficult to determine precise start and stop times for this outburst.
In addition, there were no SMC X-2 observations between approximately
March 6 to April 1 (MJD
51609 to 51635).  To obtain a crude characterization of the outburst in
a quantitative way we fitted a model consisting of a constant flux plus
a symmetric triangular profile outburst to the ASM light curve. This fit
yields an outburst starting on MJD 51555.6 $\pm$ 12.2 (2000 January 13)
and reaching a peak 47 $\pm$ 6 days later (February 28). However, 
since the gap in the ASM observations started
at close to the fitted peak time, the time of real peak
flux may actually have been later than this.
The fitted maximum count rate is 1.0 $\pm$ 0.25 counts s$^{-1}$
which, using the spectral parameters determined with the PCA (Section
3.2) corresponds to absorbed and unabsorbed 2 $-$ 10 keV luminosities of 
\sqig 1.5 
and \sqig 1.7 $\times$ 10$^{38}$
ergs s$^{-1}$ respectively at 65 kpc. The peak flux during this
outburst was thus at least comparable to, and may even have been
higher than, the \sqig 0.8 $\times$ 10$^{38}$ (d/65 kpc)$^2$
ergs s$^{-1}$ reported by Clark et al. (1978).
Although the end of the outburst based on the
fit of this simple model corresponds to roughly MJD 51649 (2000 April 15)
the PCA observations (see next section) show activity after this time.

In an attempt to determine the orbital period of \src\ we searched for
periodic modulation in the ASM light curve by calculating power spectra
but find no modulation at an amplitude $>$ 0.1 ASM counts s$^{-1}$ (\sqig1 mCrab)
for periods longer than 0.5 days.

\subsection{PCA}

 From the slews made over the location of
\src\ in observations \#1 and \#2 we obtain a source location of
R.A. = 00$^h$54$^m$53$^s$, decl. = $-$73\degrees 38\arcmin\ with a 90\% confidence error circle of
3\arcmin\ radius. This is consistent with the location of SMC X-2
at R.A. = 00$^h$54$^m$33$^s$.4, decl. = $-$73\degrees 41\arcmin\ 04\arcsec
(J2000; Murdin et al. 1979).

During the pointed observations we searched for pulsations by calculating
power spectra from the PCA light curve and in observations \#2 and \#3
pulsations were very strongly detected at periods of approximately 2.37
seconds. In order to refine our estimates of the period we performed
a pulse time arrival analysis. For each observation a template was
constructed by folding the light curve on the period found from the
power spectrum. The light curves were then divided into sections each
of approximately equal duration with breaks at the ends of ``good time''
intervals and these sections of the light curve were also folded on this
period. The phase shifts between the folded profiles and the template
were then calculated by cross-correlation. A linear fit was made to the
phase shifts found and this was used to re-determine the period and the
associated one $\sigma$ errors.  In this way pulse periods of 2.371532 $\pm$ 0.000002 s 
and  2.371861 $\pm$ 0.000003 s were measured from observations \#2
and \#3 respectively. The PCA light curve from observation \#3 folded
on the pulse period in four energy bands is plotted in Figure 2.

For the two pointed observations when the source was detected
(\#2 and \#3) we fit
the background subtracted spectrum between 3 to 15 keV with an absorbed
power-law model.  The results of these fits are given in Table 3.
The unabsorbed 2 -- 10 keV luminosities implied by these fits are 
\sqig 3.3 $\times$ 10$^{37}$ (d/65 kpc)$^2$ ergs s$^{-1}$
and
\sqig 2.7 $\times$ 10$^{37}$ (d/65 kpc)$^2$ ergs s$^{-1}$
respectively.

In the last two observations made on May 2-3 (\#4) and May 5 (\#5)
no
pulsations and no significant flux were detected.
Given the large field of view of the PCA and the possibility
of contaminating sources the most stringent upper limit
on the source flux probably comes from the lack of detection
of any pulsations rather than the nominally background
subtracted spectrum.
Based
on the lack of pulsations, and scaling from the two detections
in observations \#2 and \#3,
we place upper limits on the source flux of \sqig0.38 $\times$
10$^{-11}$ ergs cm$^{-2}$ s$^{-1}$ for observation \#4
and \sqig0.6 $\times$
10$^{-11}$ ergs cm$^{-2}$ s$^{-1}$ for observation \#5.
These flux limits corresponds to luminosities of $<$ 
\sqig 1.8 $\times$ 10$^{36}$ (d/65 kpc)$^2$ ergs s$^{-1}$
and
$<$ 
\sqig 2.9 $\times$ 10$^{36}$ (d/65 kpc)$^2$ ergs s$^{-1}$
respectively.

\subsection{Optical}

 From Table 2 it can be seen that the proposed optical counterpart of
\src\ (Star B) was approximately 0.5 magnitudes fainter in April 2000
than in 1978 and somewhat redder, which is likely indicative of less
circumstellar disk emission. However, when Murdin et al. performed their
optical observations on 1978 June 15 the source had already faded below
the detectable X-ray level: Clark et al (1979) had observed \src\ again
on 1977 December 7-15 - just 2 months after their original detection
and failed to detect it with a 3$\sigma$ upper limit of 1.0 $\times$
10$^{37}$ ergs s$^{-1}$.  In contrast we measured a flux of \sqig 3.3
$\times$ 10$^{37}$ ergs s$^{-1}$ just two days before our optical
observations. The X-ray and optical brightness of SMC X-2 may thus
not be correlated in a simple fashion.  For neither the 1978 or the
2000 outburst do we have optical photometry obtained at the peak of the
outbursts and it is possible that the optical counterpart may have been
brighter during the peak.

\section{Discussion}

Given the large number of X-ray pulsars in the SMC can we be certain
that the new pulsar that we have detected is indeed SMC X-2?
The pulse period is rather short and we can make use
of the result of Stella,
White \& Rosner (1986) who
noted an anti-correlation between pulse period
and maximum X-ray luminosity. SMC X-2 at its peak observed with SAS-3 
and the RXTE ASM was
rather bright at $>$ \sqig 10$^{38}$ ergs s$^{-1}$ which, together
with the pulse period that we measure, would be consistent
with the relationship found by Stella et al.

The vicinity of SMC X-2 may also be less densely populated with active
pulsars than the region around SMC X-3. After emission had been detected
from the vicinity of SMC X-3 with RXTE subsequent analysis revealed the
presence of three pulsars (Corbet et al. 1998), while ASCA observations
showed that none of these were in fact SMC X-3 itself. However, in
the case of SMC X-2, we only detect one pulsar and, in addition, its
location is consistent at the \sqig 3 arc minute level with SMC X-2. The
brightening observed in the ASM light curve for SMC X-2 is also consistent
with the source being not more than a few arcminutes from SMC X-2.
Further evidence still for the identification of the 2.37s pulsar with
SMC X-2 comes from the detection of this source by Torii et al. (2000)
with the Gas Imaging Spectrometer experiments on board ASCA.
The GIS is an imaging instrument which gives source locations with 90\%
errors of 1 to 2 arc minutes radius (Tashiro et al. 1995). In the ASCA GIS
survey of the SMC (Yokogawa et al. 2000) 12 X-ray pulsars and 8 candidate
X-ray pulsars were identified. We thus obtain an approximate estimate of
the chance of finding any X-ray pulsar or candidate within 2 arc minutes
of SMC X-2 in a snap-shot survey of a few percent.\
Another independent estimate of the probability that the source that we
have observed is SMC X-2 comes from the serendipitous RXTE slew detection
(observation \#1). The slews across the SMC that occur during routine
RXTE observations provide a monitor for bright new sources within this
entire region.  There are currently only five sources known in the SMC
with luminosities \sqig10$^{38}$ ergs s$^{-1}$: SMC X-1, SMC X-2,
SMC X-3, H0107-750, and the 31 second pulsar XTE J0111.2-7317 (Chakrabarty
et al. 1998).
The probability of
finding a new source with a comparable brightness at a position consistent
with one of these known bright sources within the approximately
2\degrees\ $\times$ 2\degrees\ extent of the SMC is less than 1\%.

Together, the comparable luminosity observed with RXTE and SAS-3 and
the positional coincidence at the few arc minute level, strongly suggest
that the source we have observed is indeed SMC X-2. The evidence would
be strengthened further still if the position of the 2.37 s pulsar could
be localized more precisely or evidence found for pulsations at this
period in the archival SAS-3 or HEAO-1 observations.

The outburst we have observed with RXTE could be due to either changes
in the circumstellar envelope of the Be star, associated with orbital
variability, or both of these effects.  We note that no other outburst
of a similar size is apparent in the ASM light curve which would argue
for a short lived change in the circumstellar envelope.  If \src\ has
properties consistent with the general correlation between orbital and
pulse periods exhibited by most Be/neutron star binaries (Corbet 1986)
then an orbital period of order 20 days to within a factor \sqig2 might
be expected. However, no evidence for any periodic variability is found
in the ASM light curve.

During the course
of our observations the pulse period of \src\ increased and was longer
still at 2.37230 $\pm$ 0.00004 s when ASCA observed on April 25-26 (Torii
et al. 2000). This change in pulse period may be due to orbital motion
and would correspond to a velocity change of \sqig 100 km s$^{-1}$.
While there are currently too few pulse period measurements
to place any strong constraints on the orbital parameters,
if \src\ returns again to an active state it is possible that additional
observations may yield the orbital period, either from monitoring the
pulse period or from periodic outbursts that the ASM could observe.

The pulse period that we find for \src\ is among the shortest
currently known for high mass X-ray binaries (e.g. Bildsten et
al. 1997) especially for those systems which contain Be
star primaries. As noted above, this short pulse period is consistent
with the high source luminosity. SMC X-3, which also exhibited
a high maximum luminosity, may also be expected to have a relatively
short pulse period.
Previously Clark et al. (1978, 1979)
argued that the high luminosities of SMC X-1, X-2 and X-3
implied that the luminosity distribution of the SMC X-ray
sources is shifted toward higher luminosities compared to
Galactic systems. However, now that the transient X-ray
pulsar population of the SMC is known to be so large
this shift to higher luminosities is called into question
(see e.g. Haberl \& Sasaki 2000)
as the earlier observations were only detecting the
high-luminosity tip of the large SMC X-ray pulsar population.
Extensive studies of these objects down to low flux
levels are required to fully establish their overall properties
as a class.

\acknowledgments
This paper made use of quick look data provided by the RXTE ASM team at
MIT and GSFC. We thank our colleagues in the RXTE team at GSFC
and MIT for useful
discussions on several aspects of RXTE data analysis.


\pagebreak
\noindent
{\large\bf Figure Captions}

\figcaption[asm_lc.ps]
{Light curve of \src\ obtained with the RXTE All Sky Monitor. This was
derived from the standard
daily averaged light curve by a procedure that combines weighted
rebinning by a factor of two followed by a smoothing function that replaces
each point with half its original value plus one quarter of the
value of each
neighboring point. This function was applied
twice to the original light curve.}

\figcaption[epulse_fold.ps]
{PCA light curve of \src\ obtained on 2000 April 22 - 23 
(observation \#3) folded on the
pulse period of 2.371861s. The light curve was obtained from the three
PCUs that were operational during this observation and
the background level has been subtracted for each energy band.
The phase convention is arbitrary.
}


\begin{figure}
\plotone{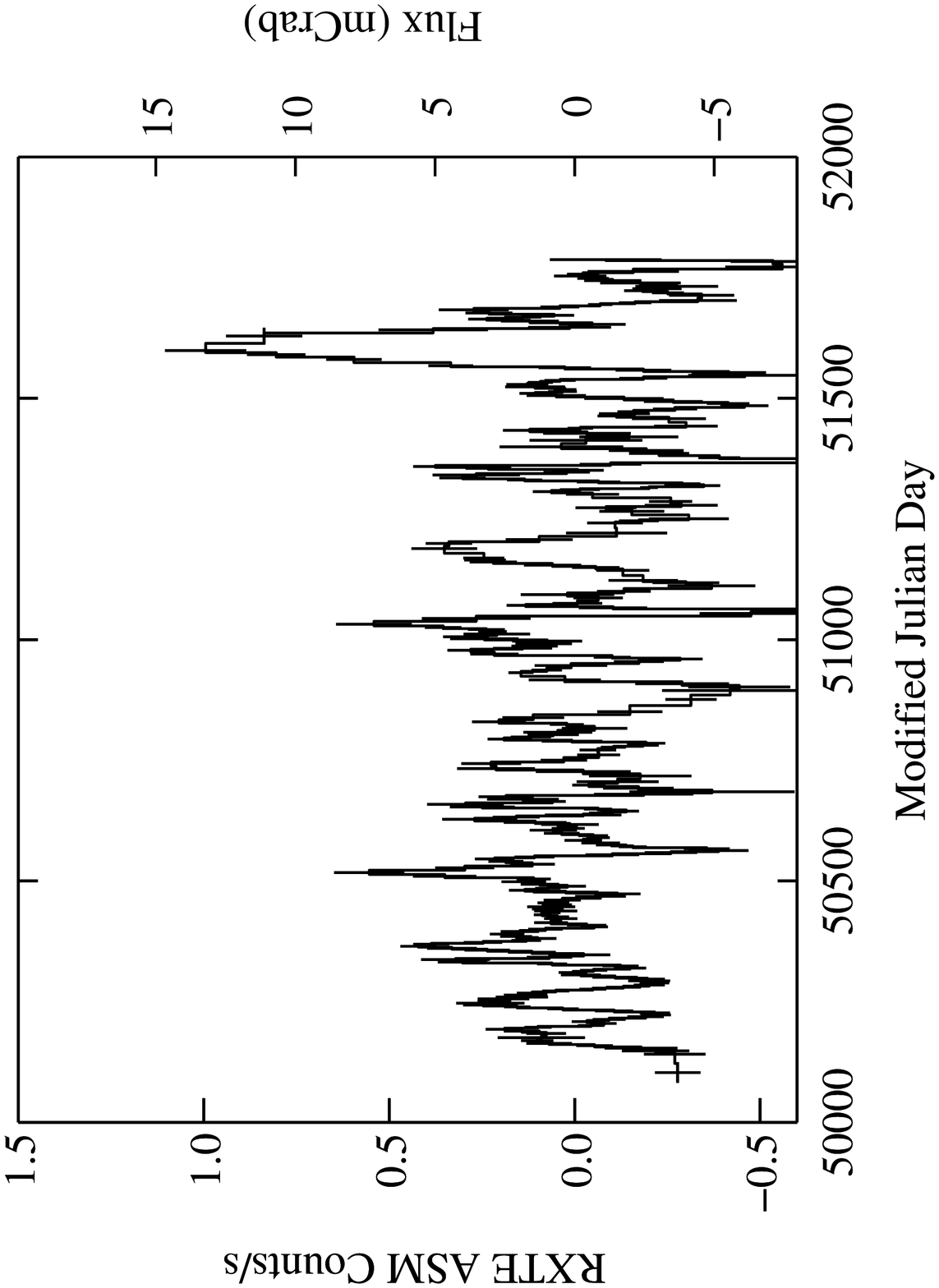}
\end{figure}


\begin{figure}
\plotone{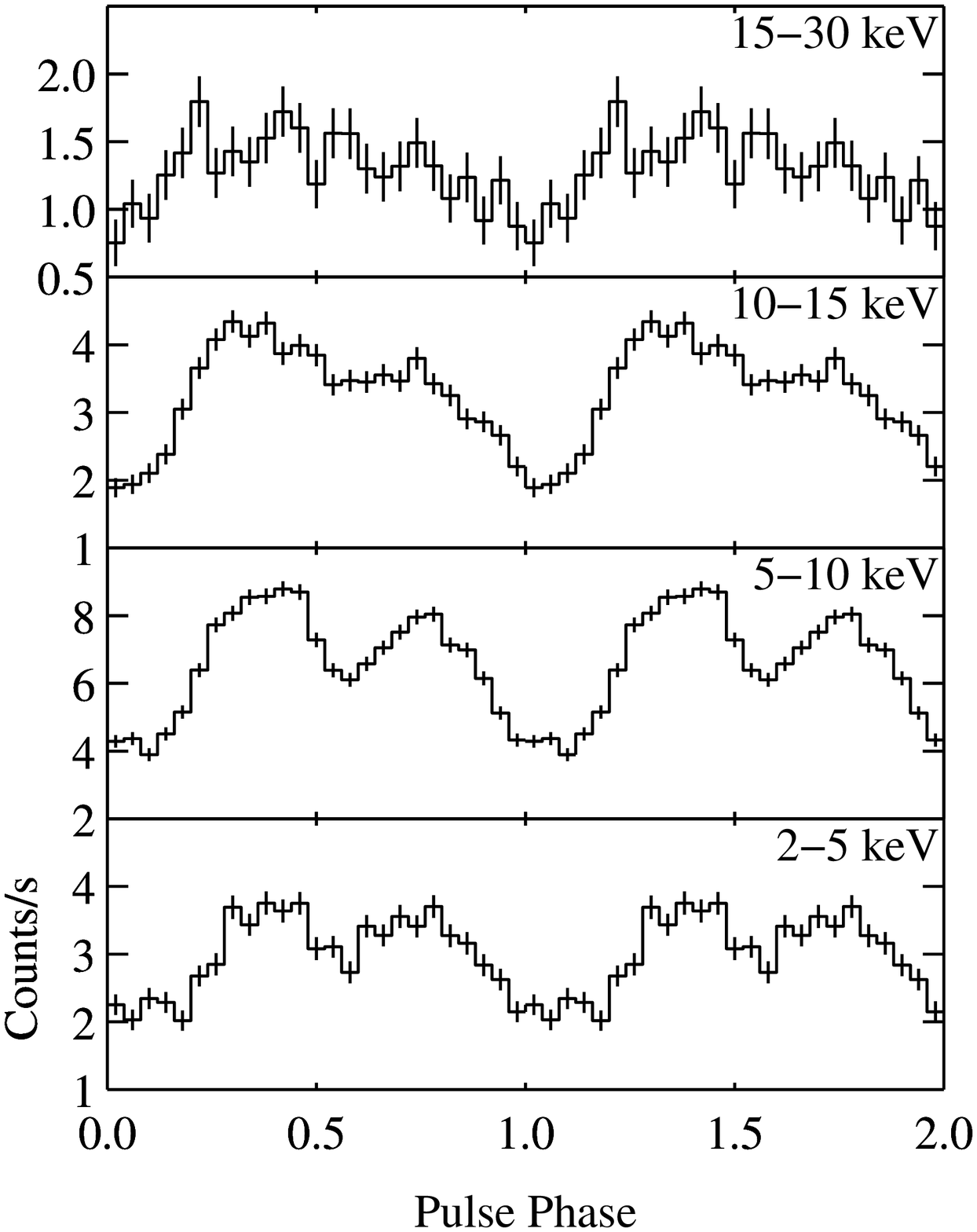}
\end{figure}

\pagebreak

\begin{table}
\begin{center}
\caption{RXTE PCA Observations of \src}
\begin{tabular}{clccccc}
\tableline
\tableline
Observation & Dates  & Exposure & Type & PCUs & Detection? & PCA\\
Number& &  (ks) & & Operating & &Count Rate\\
\tableline
1 & 2000 Apr 9 &  0.02 & Serendipitous Slew & 2 & Yes & 8 \\
2 & 2000 Apr 12  & 1.4 & Slews + Stares & 4 & Yes & 5.2\\
3 & 2000 Apr 22-23 & 7.1 & Stare & 3 & Yes & 3.2 \\ 
4 & 2000 May 2-3 & 6.3 & Stare & 4/5 & No & $<$ 0.35\\
5 & 2000 May 5  & 2.0 & Stare & 4 & No & $<$ 0.5 \\
\tableline
\end{tabular}
\end{center}
Notes: PCA count rate is counts s$^{-1}$ PCU$^{-1}$ (2 -- 10 keV) and is
approximate for observation \#1. The two upper limits come from the
lack of pulsations (see Section 3.2).
\end{table}

\begin{table}
\begin{center}
\caption{Optical Photometry}
\begin{tabular}{lccccccc}
\tableline
\tableline

        &      &  2000 Apr 14 &         & &              &  1978 June 15 &\\
\tableline
       &  B    &   V    &   R$_C$    & &          B    &  V     &   R$_C$ \\
\tableline
Star A & 15.01 &  14.88 &  14.83 & &         15.0  &  15.2  &    -- \\
Star B & 16.47 &  16.38 &  16.41 & &         15.7  &  16.0  &    -- \\
\tableline
\end{tabular}
\end{center}
Notes: The 1978 photometry is from Murdin et al. (1979).
Star B is the proposed optical counterpart of \src.
\end{table}

\begin{table}
\begin{center}
\caption{Results of Spectral Fits to PCA Data}
\begin{tabular}{ccccc}
\tableline
\tableline
Observation & $\alpha$ & N$_H$ & $\chi^2_\nu$ & 2-10 keV Flux  \\
Number &  & ($\times$ 10$^{22}$) &  & (ergs cm$^{-2}$ s$^{-1}$)\\
\tableline
2 & 0.70 $\pm$ 0.04 & 1.6 $\pm$ 0.8 & 1.52 &  6.93 $\times$ 10$^{-11}$\\
3 & 1.00 $\pm$ 0.03 & 3.3 $\pm$ 0.6 & 1.12 & 5.71 $\times$ 10$^{-11}$\\
\tableline
\end{tabular}
\end{center}
Notes: The flux is the unabsorbed flux from the model fits.
$\alpha$ is the photon index of the fitted power law.
\end{table}

\end{document}